\documentstyle[multicol,aps,prl]{revtex}
 
\input{epsf}
 
\begin{document}

\draft 

\title{Stochastic field theory for a Dirac particle
  propagating in gauge field disorder}
 
\author{T.~Guhr, T. Wilke, and H.A.~Weidenm\"uller}

\address{Max--Planck--Institut f{\"u}r Kernphysik, Postfach 103980, 
  69029 Heidelberg, Germany}
 
\date{\today}
 
\maketitle
 
\begin{abstract}
  Recent theoretical and numerical developments show analogies between
  quantum chromodynamics (QCD) and disordered systems in condensed
  matter physics.  We study the spectral fluctuations of a Dirac
  particle propagating in a finite four dimensional box in the
  presence of gauge fields. We construct a model which combines
  Efetov's approach to disordered systems with the principles of
  chiral symmetry and QCD. To this end, the gauge fields are replaced
  with a stochastic white noise potential, the gauge field disorder.
  Effective supersymmetric non--linear $\sigma$--models are obtained.
  Spontaneous breaking of supersymmetry is found.  We rigorously
  derive the equivalent of the Thouless energy in QCD.  Connections to
  other low--energy effective theories, in particular the
  Nambu--Jona-Lasinio model and chiral perturbation theory, are found.
\end{abstract}
 
 
\begin{multicols}{2}
 
\narrowtext
 
The propagation of an electron in a finite sample, such as a piece of
wire, becomes diffusive due to multiple scattering at the impurities.
The ensemble of the impurities is referred to as disorder. The
diffusion constant of this process determines, together with the size
of the sample, a universal energy scale, the Thouless energy.  This,
in turn, measured in units of the single particle mean level spacing
yields the dimensionless conductance of the wire, see the review in
Ref.~\cite{review}.  Recently, it has been argued that the
Gell-Mann--Oakes--Renner (GOR) relation has for QCD an analogous
meaning \cite{Jan98a,Osb98a,Osb98b}, implying the existence of a
Thouless energy in QCD.  Since the Thouless energy sets the scale
within which the fluctuation properties are fully of Wigner--Dyson
type, it can be found by analyzing the spectral statistics.  Indeed,
this scale was identified in data of lattice gauge calculations
\cite{Ber98b,Guh99}.
 
\begin{figure}[ht]  
\centerline{\epsfxsize=0.8\columnwidth 
            \epsffile{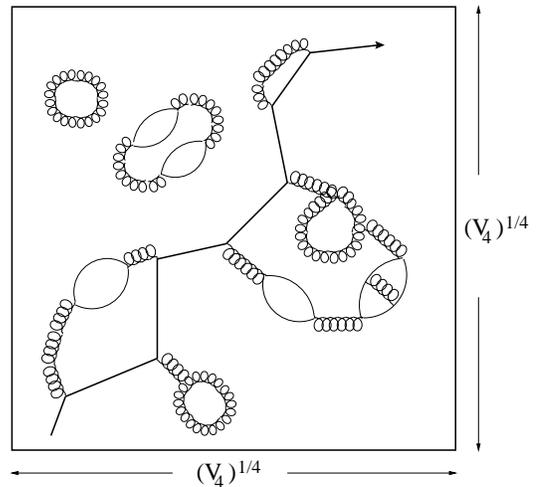}}
\caption{Propagation of a quark (thick solid line) through the 
  QCD vacuum in a box of volume $V_4$. The vacuum fluctuations involve
  quarks (thin solid lines) and gluons (twisted lines).  The
  propagating quark is scattered at these vacuum fluctuations. 
\label{fig1}}
\end{figure}

Our goal is the description of the diffusion process for QCD by a
stochastic field theory.  To this end, we merge Efetov's
supersymmetric approach to disordered systems~\cite{Efe83} with the
principles of QCD and chiral symmetry.  This extends and complements
the semi--classical reasoning of Refs.~\cite{Jan98a,Osb98a,Osb98b}.
We will not use formal analogies to the GOR relation as in Refs.
\cite{Jan98a,Osb98a,Osb98b}, rather we will derive the Thouless
energy as a natural result from our stochastic model.  Anticipating
our analytical findings, we draw an intuitive picture in
Fig.~\ref{fig1}.  Due to spontaneous breaking of chiral symmetry, the
gluonic gauge fields in the Yang--Mills action generate closed gluon
and fermion loops in the QCD vacuum. We may view these vacuum
fluctuations as ``gauge field disorder'' for the propagation of a
constituent quark or, more generally, of a Dirac particle in a four
dimensional box of finite volume $V_4$. The multiple scattering at the
vacuum fluctuations renders the motion of the quark diffusive. In this
way, the pion decay constant and the constituent quark mass are
generated.
 
For our stochastic model, we introduce the Euclidean Dirac operator in
four dimensions
\begin{equation}
iD[u]=\left[
\begin{array}{cc} 0 & \sigma_\mu\partial_\mu-ig\sigma_\mu u_\mu(x) \\ 
  \left(\sigma_\mu\partial_\mu-ig\sigma_\mu u_\mu(x)\right)^\dagger & 0 
\end{array}
\right] \ ,
\label{eq1}
\end{equation}
where $\sigma_i$, $i=1,2,3$ are the Pauli matrices,
$\sigma_4=\openone_2$, and $g$ is the gauge coupling.  The complex
fields $u_\mu(x)$ describe the gauge field disorder. We choose them as
stochastic white noise potentials with moments $\langle
u_\mu(x)\rangle_u=0$ and $\langle u_\mu(x)u_\nu(y)\rangle_u
=\gamma\delta_{\mu\nu}\delta^{(4)}(x-y)$.  The strength constant
$\gamma$ will be determined later.  Thus, the average of a quantity
${\cal R}[u]$ is given by
\begin{equation}
\langle{\cal R}[u]\rangle_u
  =\int d[u]{\cal R}[u] \exp\left(-\frac{1}{\gamma}
                 \int d^4x u_\mu(x) u_\mu^*(x)\right) \ .
\label{eq2}
\end{equation}
This replaces in our model the average over the non--Abelian gauge
fields in the Yang--Mills action.
Thus, we abandon gauge invariance, as can be seen
from Eq.~(\ref{eq2}). We are free to do so because we are exclusively
interested in stochastic features of the spectrum. 
The structure of the original gauge group enters in the model only
through the assumption that the Dirac operator is proportional
to the unit matrix $\openone_{N_c}$ in color space. This is
fully consistent with standard models of QCD~\cite{Kle92,Sch98}.

The eigenvalue equation $iD[u]\psi_i=\lambda_i[u]\psi_i$ defines for
each realization of the disorder potential $u(x)$ a spectrum
$\{\lambda_i[u]\}$.  We wish to discuss the spectral 
correlation functions of $k$ eigenvalues $\lambda_p, p=1,\ldots,k$
\begin{equation}
\widehat{R}_k(\lambda_1,\ldots,\lambda_k) 
 =\frac{1}{\pi^k}
 \left\langle\prod_{p=1}^k{\rm tr}
 \frac{1}{\lambda_p-iD[u]}\right\rangle_u \, . 
\label{eq4}
\end{equation}
The energies are given proper imaginary increments to remove the
singularities on the real axis.  The functions~(\ref{eq4}) can be
obtained as derivatives of a generating function with respect to
external sources $J_1,\ldots,J_k$,
\begin{equation}
\widehat{R}_k(\mbox{\boldmath$\lambda$\unboldmath})
= \left.
\frac{1}{(2\pi)^k}\frac{\partial^k}{\prod_{p=1}^{k}\partial J_p}
Z_k(\mbox{\boldmath$\lambda$\unboldmath}+{\bf J})\right|_{{\bf J}=0} \ ,
\label{eq5}
\end{equation}
where we have introduced $\mbox{\boldmath$\lambda$\unboldmath}={\rm
diag}(\lambda_1,\ldots,\lambda_k)\otimes 1_2$ and ${\bf J}={\rm
diag}(J_1,\ldots,J_k)\otimes \hat{k}$ with $\hat{k}={\rm
diag}(-1,+1)$.  In the supersymmetry method, the generating or
partition function is expressed as a functional integral over the
$4\cdot2k$--component and space time dependent superfields $\chi(x)$
through
\begin{eqnarray}
\lefteqn{
  Z_k(\mbox{\boldmath$\lambda$\unboldmath}+{\bf J}) = \left\langle
  {\rm detg}D[u;\mbox{\boldmath$\lambda$\unboldmath}+{\bf
  J}]\right\rangle_u }
\nonumber\\ &=& 
\left\langle\int d[\chi]
\exp\left(i\int d^4x\chi^\dagger(x) 
D[u;\mbox{\boldmath$\lambda$\unboldmath}+{\bf J}]\chi(x)\right)\right\rangle_u
\label{eq8}
\end{eqnarray}
with detg denoting the superdeterminant. Here, we have defined
\begin{equation}
D[u;\mbox{\boldmath$\lambda$\unboldmath}+{\bf J}] =
\left(\mbox{\boldmath$\lambda$\unboldmath}+{\bf
J}\right)\otimes\openone_4 - \openone_{2k}\otimes i D[u] \ .
\label{eq6}
\end{equation}
We suppress a Kronecker product with $\openone_{N_c}$.  The entries of
$\mbox{\boldmath$\lambda$\unboldmath}+{\bf J}$ can be viewed as
virtual valence quark masses.  Due to the chiral structure of the
Dirac operator the superfields $\chi(x)$ can be decomposed in left and
right handed parts $\chi_{L,R}(x)=P_{L,R}\chi(x)$, where $P_{L,R}$ are
the left and right handed projectors acting on the Dirac spinors.  The
disorder average according to Eq.~(\ref{eq2}) can now be performed and
yields a field theory quartic in $\chi(x)$.  Similar to the steps
taken in Ref.~\cite{Guh97}, it can be decoupled via a
Hubbard--Stratonovich transformation by introducing $2k\times2k$
complex supermatrix fields $\sigma(x)$ such that the fields $\chi(x)$
can be integrated out,
\begin{equation}
Z_k(\mbox{\boldmath$\lambda$\unboldmath}+{\bf J}) 
=2^{2k^2}\int d[\sigma]d[\sigma^\dagger]
\exp\left(-F[\sigma,\sigma^\dagger; 
\mbox{\boldmath$\lambda$\unboldmath}+{\bf J}]\right) \ .
\label{eq9}
\end{equation}
Both, $\sigma(x)$ and its complex conjugate $\sigma^\dagger(x)$ are
independent variables. This reflects that $\chi_L(x)$ and $\chi_R(x)$
are independent degrees of freedom in the chiral symmetric phase.  The
free energy reads
\begin{eqnarray}
F[\sigma,\sigma^\dagger;\mbox{\boldmath$\lambda$\unboldmath}+{\bf J}]&=&
{\rm trg}\int d^4x\Big(\frac{1}{4\gamma g^2}\sigma(x)\sigma^\dagger(x)
\nonumber\\&&
+{\rm tr}\log
D[\sigma,\sigma^\dagger;\mbox{\boldmath$\lambda$\unboldmath}+{\bf
J}]\Big) \ .
\label{eq10}
\end{eqnarray}
The trace ${\rm tr}={\rm tr}_{\rm Dirac}{\rm tr}_{\rm color}$ in front
of the logarithm is taken over both Dirac and color indices, the
supertrace trg is defined in the space of supermatrices.  The
supermatrix analogue of the Dirac operator is given by
\begin{eqnarray}
D[\sigma,\sigma^\dagger;\mbox{\boldmath$\lambda$\unboldmath} 
+{\bf J}]&=&
\sigma(x)\otimes P_R+\sigma^\dagger(x)\otimes P_L 
\nonumber\\&&
+\left(\mbox{\boldmath$\lambda$\unboldmath}+{\bf
J}\right)\otimes\openone_4-
\openone_{2k}\otimes(i\!\not\!\partial) \ ,
\label{eq11}
\end{eqnarray}
and describes a Dirac particle coupled to the supermatrix fields
$\sigma(x)$.  An important observation from Eq.~(\ref{eq11}) is that
$\sigma(x)$ is coupled only to the right handed modes of the Dirac
spinors and $\sigma^\dagger(x)$ only to the left handed modes.  This
implies chiral symmetry, since $\sigma(x)$ and $\sigma^\dagger(x)$ are
independent variables.
 
In contrast to supersymmetry in standard high energy physics, neither
$\chi(x)$ nor $\sigma(x)$ directly represent physical particles, such
as quarks or mesons. Rather, these fields are the degrees of
freedom in the stochastic model.  Nevertheless, the functionals
defined in Eqs.~(\ref{eq8}) and (\ref{eq10}) posses certain symmetries
which can be interpreted as the supersymmetric version of particle
symmetries known from QCD.  The Lagrangian in Eq.~(\ref{eq8}) is
invariant under the chiral transformations
\begin{eqnarray}
\chi_{L}(x) &\longrightarrow& u_{L} \chi_{L}(x)
 \ ,\quad u_L {u_L}^\dagger={u_L}^\dagger u_L=\openone_{2k}
\nonumber\\
\chi_{R}(x) &\longrightarrow& u_{R} \chi_{R}(x)
 \ ,\quad u_R {u_R}^\dagger={u_R}^\dagger u_R=\openone_{2k} \ ,
\label{eq12}
\end{eqnarray}
with $u_{L,R} \in {\rm SU}(k/k)$. We notice that ${\rm U}(k/k)$ is
likely to lead to axial anomalies which we do not discuss here.  
The corresponding supersymmetric version of chiral symmetry is 
${\rm SU}_L(k/k)\otimes{\rm SU}_R(k/k)$.  The generating
functional~(\ref{eq9}) exhibits again chiral invariance
\begin{eqnarray}
\sigma(x) \to u_L\sigma(x)u_R
\quad{\rm and}\quad
\sigma^\dagger(x) \to u_R^\dagger\sigma^\dagger(x) u_L^\dagger \ ,
\label{eq14}
\end{eqnarray}
if the explicit symmetry breaking is turned off,
i.e.~$\mbox{\boldmath$\lambda$\unboldmath}+{\bf J}=0$.  This is not
true for non--zero values of the virtual quark masses,
$\mbox{\boldmath$\lambda$\unboldmath}+{\bf J}\neq0$ which explicitly
breaks chiral symmetry. However, there is a remnant symmetry.
Consider the case of $k$ degenerate imaginary valence quark masses,
$\lambda_1=\ldots=\lambda_k=\lambda$,
i.e.~$\mbox{\boldmath$\lambda$\unboldmath}+{\bf
J}=\lambda\openone_{2k}$, where the sources are set to zero.  Then the
functional is invariant under the transformation
\begin{eqnarray}
\sigma(x) \to u\sigma(x)u^\dagger
\quad{\rm and}\quad
\sigma^\dagger(x) \to u\sigma^\dagger(x) u^\dagger \ ,
\label{eq15}
\end{eqnarray}
with $u \in {\rm SU}(k/k)$.
 
We observe a close relation between the one--point function in the
microscopic regime (near zero virtuality) and the two--point function
in the bulk (far away from zero virtuality).  In both cases, the
symmetry breaking is controlled by a single parameter, although the
symmetry groups are different. In the bulk, the symmetry is broken by
non--degenerate eigenvalues $\lambda_1$ and $\lambda_2$, and the
symmetry breaking parameter is $\omega=\lambda_2-\lambda_1$. In the
microscopic region, the symmetry is already broken by a non--zero
eigenvalue $\lambda=\lambda_1$.
 
In disordered systems, to make analytical progress, one considers the
field theory via a saddle point approximation in the limit of weak
disorder, i.e.~for a relatively low density of impurities.  This
corresponds to a Born approximation of the self--energy where diagrams
with crossed interaction lines are neglected \cite{Efe83}.  Here, we
proceed similarly by a saddle point approximation of the partition
function~(\ref{eq9}) in the limit of weak gauge field disorder. This
coincides with a $1/N_c$ expansion, which is a standard asymptotic
limit in many QCD models. These two limits coincide because for
$N_c\to\infty$ and $N_cg^2={\rm const}$ \cite{tHo74,Wit79} the
relevant diagrams contributing in leading order to the self--energy
within our model are analogous to the ones in disordered systems.
This shows the equivalence of the $1/N_c$ expansion and the weak
disorder limit.  {}From the saddle point condition $\delta
F[\sigma,\sigma^\dagger; \mbox{\boldmath$\lambda$\unboldmath}+{\bf
  J}]=0$, we find the self--consistent matrix equation in momentum
representation
\begin{eqnarray}
  {\Sigma_0}^\dagger&=&-8\gamma N_cg^2
  \left({\Sigma_0}^\dagger+\mbox{\boldmath$\Xi$\unboldmath}\right)
  \nonumber\\&&\hspace{-1cm}\times\!\!\int\limits^{\Lambda_{\rm cut}}
  \frac{d^4p}{(2\pi)^4}
  \Big[\Big(\Sigma_0+\mbox{\boldmath$\Xi$\unboldmath}\Big)
  \Big({\Sigma_0}^\dagger+\mbox{\boldmath$\Xi$\unboldmath}\Big)
  -\openone_{2k}\otimes\openone_4 p^2\Big]^{-1}
\label{eq17}
\end{eqnarray}
with
$\mbox{\boldmath$\Xi$\unboldmath}=(\mbox{\boldmath$\lambda$\unboldmath}
+{\bf J})\otimes\openone_4$ and $\Sigma_0=\sigma_0\otimes
P_R+\sigma_0^\dagger\otimes P_L$, where $\sigma_0$ and
$\sigma_0^\dagger$ are the fields at the saddle point.  A cutoff
$\Lambda_{\rm cut}$ has to be introduced to regularize the divergent
momentum integration.  For scalar fields, the saddle point equation
(\ref{eq17}) coincides with the gap equation of the
Nambu--Jona-Lasinio (NJL) model \cite{Nam61}. The NJL model is a
theory for chiral fermions coupled via a four--point interaction.  The
gap equation describes the self--energy of a quark calculated in the
self--consistent Hartree approximation obtained from a $1/N_c$
expansion \cite{Kle92}.  For vanishing virtual quark masses,
$\mbox{\boldmath$\lambda$\unboldmath}+{\bf J}=0$, there exists only a
non--trivial solution, indicating chiral symmetry breaking, if the
coupling exceeds a critical value $\gamma g^2>2\pi^2/N_c\Lambda_{\rm
cut}^2$. For non--vanishing masses, however, one complex solution
always exists, which we denote by $M^*_\lambda$.  Since the spectral
density per four--volume is given by $\nu(\lambda)={\rm
Im}\widehat{R}_1(\lambda)$, see Eq.~(\ref{eq4}), the value of the
constant $\gamma$ in Eq.~(\ref{eq2}) is now fixed. {}From
Eq.~(\ref{eq17}) we find $\gamma={\rm Im}M^*_\lambda/(\pi
g^2\nu(\lambda))$.
 
In contrast to the situation in disordered systems, we have to
distinguish the microscopic and the bulk region.  In the latter, the
symmetries of the saddle point solution are still the same as given in
Eq.~(\ref{eq15}).  A solution of the saddle point equation is
\begin{equation}
\sigma_0={\sigma_0}^\dagger
=i{\rm Im}M^*_\lambda L \ , 
\label{eq17b}
\end{equation}
where $L=\openone_4\otimes\hat{k}$. On the other hand, in the
microscopic region the chiral symmetry defined in Eqs.(\ref{eq12}) to
(\ref{eq14}) is not preserved at the saddle point.  The solutions read
\begin{equation}
\sigma_0(x)=iM^*\widetilde{U}_0(x) \quad{\rm and}\quad   
{\sigma_0}^\dagger(x)=iM^*{\widetilde{U}_0}^\dagger(x)
\label{eq18}  
\end{equation}  
where $\widetilde{U}_0(x)$ is $2\times 2$ unitary supermatrix field
and $M^*$ is the non--trivial scalar solution of the gap equation
(\ref{eq17}) for $\mbox{\boldmath$\lambda$\unboldmath}+{\bf J}=0$.  We
conclude that chiral symmetry is spontaneously broken such that
\begin{equation}
{\rm SU}_R(k/k)\otimes{\rm SU}_L(k/k)\longrightarrow{\rm SU}(k/k) \ .
\label{eq19}  
\end{equation}  
This is the supersymmetric analogue of spontaneous breaking of chiral
symmetry in our stochastic model.
 
By integrating out the quadratic fluctuations around the saddle
points, we obtain, first, a theory for the two--point correlations in
the bulk and, second, a theory for the one--point function in the
microscopic region.  In both cases, the zeroth order term is expanded
up to linear order in the symmetry breaking parameter, i.e.~$\omega$
in the bulk and $\lambda$ in the microscopic region, respectively.
Moreover, we restrict ourselves to fields varying slowly in
space--time, allowing for a gradient expansion of the quadratic
fluctuations around the saddle points. In the bulk we obtain for the
generating functional of the two--point function
\begin{equation}
Z_2(\lambda,\omega,{\bf J})=
 \int d\mu(Q)\exp\left(-F[Q;\lambda,\omega,{\bf J}]\right)
\label{eq21}
\end{equation}
with
\begin{eqnarray}
  F[Q;\lambda,\omega,{\bf J}]&=& \frac{\pi}{2}\nu(\lambda){\rm
    trg}\int d^4x \left({\cal
    D}(\lambda)\partial_\mu Q(x)\partial_\mu Q(x)\right.
  \nonumber\\&&
  \left.-2i\left((\omega+i\varepsilon)L+{\bf J}\right)Q(x)\right)
    \ ,
\label{eq22}
\end{eqnarray}
where $\lambda=(\lambda_1+\lambda_2)/2$. The fields $Q(x)$ are the
Goldstone modes of the saddle point manifold. They are in the coset
manifold $U(1,1/2)/(U(1/1)\otimes U(1/1))$. Here, $L$ is the symmetry
breaking matrix and $i\varepsilon$ a proper imaginary increment to
ensure convergence.  This formally extends Efetov's
result~\cite{Efe83} to four space time dimensions. In particular, the
coset spaces coincide.  Chiral symmetry is not present. This is in
agreement with its explicit breaking by the virtual current mass
$i\lambda$.  The diffusion coefficient ${\cal D}(\lambda)$, however, a
main object of our interest, must be different from disordered systems
because we start out from the Dirac equation. It is given as the
second moment
\begin{eqnarray}
{\rm tr}\int d^4y(x-y)^2
S(x,y;\lambda+i{\rm Im}M^*_\lambda)
S(y,x;\lambda-i{\rm Im}M^*_\lambda)&&
\nonumber\\&&\hspace{-5cm}=
\frac{2\pi\nu(\lambda){\cal D}(\lambda)}{({\rm Im}M^*_\lambda)^2}\,
\label{eq23}
\end{eqnarray}
where $S(x,y;\lambda-i{\rm Im}M^*_\lambda)$ is the Green function to
the mean field Dirac operator $\lambda+i{\rm
Im}M^*_\lambda-i\!\not\!\partial$.
 
{}From the competition between the kinetic and the symmetry breaking
term one can infer the Thouless energy in the bulk.  In a finite
volume $V_4$, the longest wavelength which fits inside the Euclidean
box is given by the inverse of the linear extension of the box, i.e.~
$1/(V_4)^{1/4}$.  Thus, the kinetic energy cannot be smaller than
${\cal D}(\lambda)/\sqrt{V_4}$. For smaller energies, the wavelength
exceeds the extension of the box and does not resolve details inside
it. In that case, the fields become constant and the functional is
dominated by the constant modes.  Hence, only the symmetry breaking
term gives a contribution. Then the level statistics is of
Wigner--Dyson type. If the energy exceeds this scale, the kinetic term
becomes more and more important, leading to deviations from the
Wigner--Dyson statistics. Thus, the equivalent of the Thouless energy
in the bulk is given by
\begin{equation}
\lambda_c^{\rm bulk}\sim{\cal D}(\lambda)/\sqrt{V_4} \ .
\label{eq24}
\end{equation}
 
In the chiral limit, $\lambda\to0$, the diffusion coefficient
(\ref{eq23}) in momentum representation reduces to
\begin{equation}
{\cal D}(0)=\frac{8N_c(M^*)^2}{|\langle\bar{\psi}\psi\rangle|}
\int\frac{d^4p}{(2\pi)^4}\frac{1}{\left(p^2+(M^*)^2\right)^2} \ .
\label{eq25}
\end{equation}
where $\langle\bar{\psi}\psi\rangle$ is the chiral condensate.
obtained from the Banks--Casher relation,
$\langle\bar{\psi}\psi\rangle=-\pi\lim_{V_4\to\infty}\nu(0)$.  For
this derivation, we used only the basic assumption~(\ref{eq2}) and
chiral symmetry. Low energy theorems of QCD imply that the integral in
Eq.~(\ref{eq25}) turns out to be proportional to the square of the
pion decay constant $f_\pi^2$,
\begin{equation}
f_\pi^2=4N_c(M^*)^2
\int\frac{d^4p}{(2\pi)^4}\frac{1}{\left(p^2+(M^*)^2\right)^2} \ ,
\label{eq26}
\end{equation}
for a detailed discussion see Ref.~\cite{Kle92}. With this additional
information, we arrive at
\begin{equation}
{\cal D}(0)=2f_\pi^2/|\langle\bar{\psi}\psi\rangle| \ .
\label{eq27}
\end{equation}
In Refs. \cite{Jan98a,Osb98a,Osb98b}, this formula was obtained from
the GOR relation by using formal analogies and semi--classical
considerations. Here, we gave a rigorous derivation of the diffusion
constant (\ref{eq25}) from our stochastic model.  Finally, we obtain
for the generating functional of the one--point function in the
microscopic regime
\begin{equation}
Z_1(\lambda,J)=\int
d\mu(U)\exp\left(-F[U;\lambda+J]\right) \ , 
\label{eq28}
\end{equation}
with
\begin{eqnarray}
F[U;\lambda+J]&=&
|\langle\bar{\psi}\psi\rangle|{\rm trg}\int d^4x\Big(
\frac{f_\pi^2}{|\langle\bar{\psi}\psi\rangle|}
\partial_\mu U(x)\partial_\mu U^\dagger(x)
\nonumber\\&&
-\frac{i}{2}(\lambda+J\hat{k})\left(U(x)+U^\dagger(x)\right)\Big) \ .
\label{eq29}
\end{eqnarray}
The supermatrices $U(x)$ are $2 \times 2$ unitary. Again, the
functional is dominated by the constant modes for energies smaller
than $f_\pi^2/(|\langle\bar{\psi}\psi\rangle|\sqrt{V_4})$. We conclude
that the equivalent of a Thouless energy in the microscopic region is
given by
\begin{equation}
\lambda_c^{\rm micro}\sim
{\cal D}(0)/\sqrt{V_4}=
2f_\pi^2/(|\langle\bar{\psi}\psi\rangle|\sqrt{V_4}) \ .
\label{eq30}
\end{equation}
We note that the microscopic functional reduces, up to irrelevant
numerical factors, to the partially quenched chiral perturbation
theory (pqCPT). This is consistent with the considerations of
Refs.~\cite{Osb99,Dam99}, were pqCPT was used to identify the
equivalent of the Thouless energy in the Dirac spectrum.
 
In summary, we presented a stochastic field theory by merging Efetov's
approach to disordered systems with the standard principles of QCD and
chiral symmetry. Due to the latter, we obtain different theories in
the microscopic and the bulk region which connect to the NJL model and
pqCPT.  The Thouless energy is a natural consequence of our model.  In
conclusion, we wish to make a proposal: since we now have a detailed
and, importantly, quantitative description of the spectral
fluctuations in QCD, it would be worthwhile to use this insight as an
input for lattice QCD, either by constructing effective models or by
modifying the algorithms accordingly. It would be highly desirable if
this made lattice QCD more efficient.
 
We thank R.~Baltin, S.P.~Kle\-vansky, E.~Lutz, D.~Toublan, and
J.J.M.~Verbaarschot for fruitful discussions. TG acknowledges
financial support from the Heisenberg foundation.

\end{multicols} 

\end{document}